\documentclass[useAMS,usenatbib,letterpaper]{mn2e}
\usepackage{epsfig}
\usepackage{graphicx}
\usepackage{amsmath}
\usepackage{amssymb}
\usepackage{aas_macros}

\usepackage[totalwidth=480pt, totalheight=680pt]{geometry}

\newcommand{\be}{\begin{equation}}
\newcommand{\ee}{\end{equation}}
\newcommand{\bea}{\begin{eqnarray}}
\newcommand{\eea}{\end{eqnarray}}

\newcommand {\rmd}{{\rm d}}
\renewcommand{\vec}[1]{{\bf #1}}

\title[Galaxy Zoo: Chiral correlation function of galaxy spins]
{Galaxy Zoo: Chiral correlation function of galaxy spins\thanks{This publication has been made possible by the participation
  of more than 100,000 volunteers in the Galaxy Zoo project. Their individual contributions are acknowledged at \texttt{http://www.galaxyzoo.org/Volunteers.aspx}}}
\author[An\v{z}e Slosar et al.]
{An\v{z}e Slosar$^{1,2,3}$\thanks{E-mail: anze@berkeley.edu},
Kate Land$^2$,  
Steven Bamford$^{4,5}$,
Chris Lintott$^2$, 
Dan Andreescu$^6$, \cr
Phil Murray$^7$, 
Robert Nichol$^4$, 
M. Jordan Raddick$^8$, 
Kevin Schawinski$^{9,10,2}$,  
Alex Szalay$^8$, \cr
Daniel Thomas$^4$, 
Jan Vandenberg$^8$.\\
\vspace*{-6pt} {\small \em$^{1}$ Berkeley Center for Cosmological
  Physics, Lawrence Berkeley Nat. Lab \& Phys. Dept, University of California, Berkeley, CA 94720, USA}\\
\vspace*{-6pt} {\small \em$^{2}$Astrophysics Department, University of Oxford, Oxford, OX1 3RH}\\
\vspace*{-6pt} {\small \em$^{3}$ Faculty of Mathematics \& Physics,
  University of Ljubljana, Slovenia} \\
\vspace*{-6pt} {\small \em$^4$ Institute of Cosmology and Gravitation, University of Portsmouth, Mercantile House, Hampshire Terrace, Portsmouth, PO1 2EG, UK}\\
\vspace*{-6pt} {\small \em$^5$ Centre for Astronomy and Particle Theory, University of Nottingham, University Park, Nottingham, NG7 2RD, UK}\\
\vspace*{-6pt} {\small \em$^6$ LinkLab, 4506 Graystone Ave., Bronx, NY 10471, USA}\\
\vspace*{-6pt} {\small \em$^7$ Fingerprint Digital Media, 9 Victoria Close, Newtownards, Co. Down, Northern Ireland, BT23 7GY, UK}\\
\vspace*{-6pt} {\small \em$^8$ Department of Physics and Astronomy,
  The Johns Hopkins University, Homewood Campus, Baltimore, MD 21218,
  USA}\\
\vspace*{-6pt} {\small \em$^9$ Department of Physics, Yale University,  New Haven, CT 06511, USA }\\
\vspace*{-6pt} {\small \em$^{10}$ Yale Center for Astronomy and  Astrophysics, Yale University, P.O. Box 208181, New Haven, CT 06520, USA}}

\begin{document}

\date{Accepted xxx. Received xxx; in original form xxx}

\pagerange{\pageref{firstpage}--\pageref{lastpage}} \pubyear{2006}

\maketitle

\label{firstpage}

\begin{abstract}

  Galaxy Zoo is the first study of nearby galaxies that contains
  reliable information about the spiral sense of rotation of galaxy
  arms for a sizeable number of galaxies.  We measure the correlation
  function of spin chirality (the sense in which galaxies appear to be
  spinning) of face-on spiral galaxies in angular, real and projected
  spaces.  Our results indicate a hint of positive correlation at
  separations less than $\sim$ 0.5 Mpc at a statistical significance
  of 2-3 $\sigma$. This is the first experimental evidence for chiral
  correlation of spins.  Within tidal torque theory it indicates that
  the inertia tensors of nearby galaxies are correlated. This is
  complementary to the studies of nearby spin axis correlations that
  probe the correlations of the tidal field.  Theoretical
  interpretation is made difficult by the small distances at which the
  correlations are detected, implying that substructure might play a
  significant role, and our necessary selection of face-on spiral
  galaxies, rather than a general volume-limited sample.
\end{abstract}

\begin{keywords}\end{keywords}


\section{Introduction}

Understanding the creation and evolution of the angular momentum of
dark matter halos and galaxies is a crucial building block of a
comprehensive theory of galaxy evolution. \cite{hoyle49} was first to
propose that the galaxy spin can be ascribed to the gravitational
coupling with the surrounding galaxies. This idea has been formalised
and extended in subsequent work
\citep{1969ApJ...155..393P,Dorosspin,1984ApJ...286...38W,1988MNRAS.232..339H,1996MNRAS.282..436C}
into the modern theory of the evolution of galaxy spin, known as the
tidal torque theory (TTT; see \cite{2008arXiv0808.0203S} for a
review). This theory asserts that protohalos acquire most of their
angular momentum in the early stages of their formation, from the
lowest non-vanishing contribution from the linear Lagrangian theory,
that is, a coupling of the quadrupole of the local mass distribution
to the external gravitational shear. Compared to $N$-body simulations,
theory produces qualitatively correct results, although there are
still significant discrepancies at a more quantitative
level. Moreover, it seems that at present there are no clear
theoretical directions for improving analytical models
\citep{1987ApJ...319..575B,Porciani:2001db,2005ApJ...627..647B}.

On the observational side, most of the work has been done using spiral
galaxies. These are characterised by a rotating disk of baryonic
matter. The line perpendicular to the plane of the disk determines the
axis of rotation, while the spiral arms in most galaxies encode the
sense of rotation, i.e the difference between left-hand screw and
right-hand screw sense of rotation.  For spiral galaxies seen in
projection, one can measure the observed galaxy ellipticities, which
constrain the \emph{axis} of the galaxy spin
\citep{2000ApJ...543L.107P,Lee:2001vz,Lee:2007jx,2006ApJ...640L.111T}.  This axis is known
to within two-fold degeneracy associated with the tilt of the galactic
plane with respect to the plane of the sky.  Since the vector can
point in two directions on the same axis, the ellipticities constrain
the spin vector within a total of four-fold degeneracy. Note that
chiral information, \emph{viz.} information about the actual
directions of the spin vectors as opposed to spin axis, is completely
absent in the study of galactic ellipticities.  However, this
information contains important clues about the details of the
emergence of the spin. As we will explain later in the text, the
detection of chiral correlation function implies that the local
inertia tensors must be correlated. This lends experimental support to
the theoretical expectations that the inertia and gravitational shear
tensor are correlated \citep{Porciani:2001er}.

We now have a unique tool to study the chiral properties of galaxy
spins.  Through an online project called Galaxy
Zoo\footnote{www.galaxyzoo.org} ~\citep{2008arXiv0804.4483L}, members
of the public have visually classified the morphologies and spin
orientations for the entire spectroscopic sample of the Sloan
Digital Sky Survey (SDSS from now on) \citep{2000AJ....120.1579Y} Data Release 6 (DR6)
\citep{2008ApJS..175..297A}. The data and its reduction is extensively
discussed in \cite{2008arXiv0804.4483L}.

Spiral galaxies in the Galaxy Zoo sample are classified as clockwise,
anti-clockwise or edge on. The spin direction convention used here is
such that clockwise and anti-clockwise rotations correspond to
galaxies whose arms are rotating in the sense of the letters Z and S
respectively \citep{1995MNRAS.276..327S}. For each face-on galaxy we
thus receive one bit of information corresponding to the sign of the
galaxy spin vector projected along the line of sight. It is important
to note that this information is independent of the tilt of the plane
of the galaxy.  We will refer to this one-bit information simply as
galaxy spin. By the galaxy spin vector we mean the unit vector that
defines the apparent spin of the galaxy: it is perpendicular to the
disk plane and points in the direction the right turn screw would move
if turned following the spiral arms inwards.  This quantity is
strongly correlated with the real angular momentum of the gas. The
correlation, however, is not perfect and observations show that the
angular momentum vector of the gas points in the opposite direction in
about four percent of systems \citep{1982Ap&SS..86..215P}. In turn,
there are theoretical expectations that there is a strong, but not
perfect correlation between the angular momentum vector of gas and
that of the dark matter halos hosting the galaxy
\citep{2002ApJ...576...21V}.  A detection of correlation in the galaxy
spins would therefore imply a correlation in the dark matter spin
vectors.  Conversely, a non-detection of the spin correlation can be
used to put upper limits on the correlation between angular momentum
vectors of dark matter halos.

This paper is structured as follows.  In Section \ref{sec:TTT} we
shortly review the tidal torque theory and its main results. Section
\ref{theory} will connect the correlation function $\eta$ to an
observable correlation function of spins $c$, while the Section
\ref{datamethod} will introduce our data and measurement technique. We
present our results and discuss systematics in Section
\ref{sec:results}. Finally, we discuss our results and conclude in the
last Section \ref{sec:end}.

\section{Tidal Torque Theory}
\label{sec:TTT}

The tidal torque theory derives the following expression for the
angular momentum from the 1st order linear peturbation theory in
Lagrangian space \citep{1984ApJ...286...38W,1996MNRAS.282..436C}:
\begin{equation}
  L_i(t) = a^2(t) \dot{D} \epsilon_{ijk} T_{jl}I_{lk},
\end{equation}
where $a$ is the scale factor of the Universe, $D$ is the growth
factor and $\epsilon_{ijk}$ is the Levi-Civita symbol. The local
intertia tensor $I_{ij}$ of the protohalo (the mass that will later
form the dark matter halo) in Lagrangian space is given by
\begin{equation}
  I_{ij} = \bar{\rho_o} \int_V q_i q_j \rmd^3 q,
\end{equation}
where $q_i$ are the Lagrangian coordinates around the centre of mass
of the halo and $\rho_o$ is the mean density.  The local shear tensor
$T_{ij}$ is defined by
\begin{equation}
  T_{ij} = \partial_i \partial_j \phi (\vec{q}),
\end{equation}
where $\phi$ is the gravitational potential.  In other words, the TTT
requires two components: a non-vanishing quadrupole distribution of
mass in the halo to be spun up and the cosmological tidal field. In
principle, it sounds plausible to assume that while tidal fields
between neighboring protohalos are correlated, since they are coming
from the large scale modes, the local quadrupole moments of mass
distribution are sourced due to random distribution of the local
inhomogeneities and should therefore be random.  This assumption of a
statistical isotropy of the inertia tensor, gives the following anatz
for the angular moment correlator \citep{2000ApJ...543L.107P}:
\begin{equation}
 Q_{ij} =  \left< L_i L_j |\hat{\vec{T}}\right> = \frac{1}{3}\delta_{ij} + c
  \left(\frac{1}{3} - \hat{T}_{ik}\hat{T}_{kj} \right),
\label{eq:pry}
\end{equation}
where $c$ controls the level of randomization of axial preference due
to non-linear and stochastic effects.  In fact, this ansatz has been
shown to satisfactorily explain the inclinations of axes of spiral
galaxies in vicinity of voids with $c\sim 0.7$
\citep{2006ApJ...640L.111T}. 

Some further algebra gives the probability distribution function for
spins $\vec{s}=\hat{\vec{L}}$, usually assumed to be Gaussian
\citep{2000ApJ...543L.107P}:
\begin{equation}
  P(\vec{s}|\vec{T}) = \frac{|\hat{Q}|^{-1/2}}{4\pi} \exp \left(
    -\vec{s}^{T}\cdot \hat{\vec{Q}}^{-1}\cdot \vec{s}\right).
\label{eq:prx}
\end{equation}
Using this expression it is therefore possible to calculate various
correlators of $\vec{s}$ if correlators of $\vec{T}$ are known. 

We will now consider correlation functions. The most general form of
possible correlation statistics of the galaxy spins consistent with
the homogeneity and isotropy is the spin correlation tensor defined by
\citep[see e. g.][]{Groth}
\begin{multline}
  \label{eq:tensor}
  \Xi_{ij} (r) = \left< s_i(\vec{x}) s_j(\vec{x}+\vec{r})\right> = \\
\left[\Pi(r) - \Sigma(r) \right]\hat{r}_i \hat{r}_j +  \Sigma(r)\delta_{ij}.
\end{multline}
Functions $\Pi$ and $\Sigma$ are parallel and perpendicular
correlation function. Following \cite{Porciani:2001db} we will be dealing
exclusively with the ``dot product'' correlation functions given by 
\begin{equation}
  \eta (r) = \Xi_{ii} = \left< \vec{s}(\vec{x}) \cdot \vec{s}(\vec{x}+\vec{r}) \right>,
\end{equation}
and
\begin{equation}
  \eta_2 (r) = \left< (\vec{s} (\vec{x}) \cdot \vec{s}(\vec{x}+\vec{r}))^2\right>-1/3.
\end{equation}
It is important to realize what these two quantities measure. The
first one measures if the angular momentum vectors are correlated,
while the second one measures if the axes of angular momentum vectors
are correlated. Note that is perfectly possible to have $\eta_2(r)>0$
while $\eta(r)=0$. For example, if all spins were aligned along the
$z$ axis, but with an orientation that is chosen at random from
$\hat{z}$ and $-\hat{z}$, $\eta_2=1$,
while $\eta=0$. In fact, following the ansatz of Equations
(\ref{eq:pry}) and (\ref{eq:prx}) results in vanishing $\eta(r)$ and a
finite $\eta_2(r)$ (analytical expression for which can be found in
\cite{2001ApJ...555..106L}). This is trivially seen from Equation
(\ref{eq:prx}), since $P(\vec{s}|\vec{T})=P(-\vec{s}|\vec{T})$ and is
a direct consequence of the assumption of isotropy of local moments of
inertia. To put it simply, for a fixed tidal field, averaging over
possible realisations of the inertia tensor will, in general produce a
preferred axis (determined by eigenvectors of $\vec{T}$), but not a
preferred direction. This is due to the fact that for every inertia
tensor $\vec{I}$ that produces a final angular momentum $\vec{L}$, an
equally likely mirror image inertia tensor $-\vec{I}$ will produce an
equal and opposite angular moment $-\vec{L}$. Therefore, the common
assumption that the local moments of inertia are random and
uncorrelated will, in general, produce a non-vanishing axis
correlation, but a vanishing correlation of the actual spin
vectors. The important corollary is, that a detection of chiral
correlations in the galaxy spins would directly indicate that the
moments of inertia are non-random.

\section{Correlations of projected spins}\label{theory}

\subsection{Small scale correlation function of spins}
\label{sec:small-scale-corr}

Our life as observers is complicated by the fact that the sense in
which spiral arms wind in a projected image of a spiral galaxy
measures the sign of the spin vector projected along the line of sight
rather than the spin vector itself.  We will therefore consider the
correlation function of \newcommand{\sgn}{{\rm sgn}}
\newcommand{\ts}{\tilde{s}} $\ts={\rm sgn}(\vec{s}\cdot\hat{z})$:

\begin{equation}
  c(r) = \left< \ts (\vec{x}) \ts (\vec{x}+\vec{r})) \right>,
\end{equation}
where we assumed that the radial vectors to positions $\vec{x}$ and
$\vec{x}+\vec{r}$ are parallel, i.e. the flat-sky approximation.  Note
that this only requires \emph{pairs} of galaxies to be close enough so
that the flat-sky approximation holds, rather than an entire survey
occupying a small portion of the sky. For the time being we also
neglect the difference between the dark-matter angular momentum and
the gas angular momentum.

To proceed, we note that $\eta(r)$ is determined entirely by the
one-point distribution function $P(\mu|r)$ for cosine angle $\mu=\cos
\theta$ between the two spin vectors:
\begin{equation}
  \eta(r) = \left< s_i(\vec{x}) s_j(\vec{x}+\vec{r})\right> = \left<
    \mu \right> = \int  \mu P(\mu|r) \, \rmd \mu
\end{equation}

For a given $\mu$, one will observe two galaxies with the same
orientations of spins with the probability (see beginning of the Appendix
\ref{apx})
\begin{equation}
  P_{+1} (\theta) = 1-\theta/\pi
\label{eq:plmin2}
\end{equation}
and with different orientations of spins with the probability
\begin{equation}
  P_{-1} (\theta) = 1-P_{+1} (\theta) = \theta/\pi
\label{eq:plmin}
\end{equation}

The correlation function of the spin signs is then given by
\begin{equation}
  c(r) = \frac{1}{N} \int \rmd \mu P(\mu|r) \left(P_{+1} (\theta) -
    P_{-1} (\theta)\right), 
\label{eq:2}
\end{equation}
where normalisation $N$ is in this case trivially given by:
\begin{equation}
  N= \int \rmd \mu P(\mu|r) \left(P_{+1} (\theta) + P_{-1} (\theta) \right) = 1
\label{eq:norm}
\end{equation}

When there are no correlations, $P(\mu)\,\rmd \mu= 1/2 \rmd \, \mu$
and both $\eta(r)$ and $c(r)=0$. When correlations exist, we must
specify a one point probability distribution function for
$P(\theta|r)$. We assume the following form:
\begin{equation}
  P(\theta|r) = \sin(\theta) \left(1+e(r) \cos(\theta)\right)
\label{eq:bala}
\end{equation}
Using this form, one obtains
\begin{equation}
  \eta(r) = \frac{1}{3} e(r)
\end{equation}
\begin{equation}
  c(r) = \frac{1}{4} e(r)
\end{equation}
and so
\begin{equation}
\label{eq:poncy}
  \eta(r) = \frac{4}{3} c(r)
\end{equation}
Equation (\ref{eq:poncy}) hinges on the particular form for
$P(\theta|r)$ that we chose. In practice, different forms generically
give the results that $\eta(r) = q c(r)$ with q typically between $1$
and $3/2$. 

In reality, however, we measure the correlation function only for
galaxies that are sufficiently away from the edge-on orientation not
to be classified as a face-on galaxy.  For simplicity, let us assume
that our sample contains only galaxies, whose spin vector satisfies
\begin{equation}
  s(\vec{x}) \cdot \hat{z} > \cos \alpha.
\end{equation}
In other words, galaxies that are inclined with an angle greater than
$\alpha$ with respect to the line of sight are assumed to have been
classified as edge-on.  What is the functional form for $P_{+1}
(\theta)$ in this case?  In Appendix \ref{apx}, we show that for
$\alpha > \pi/4$
\begin{equation}
  P_{+1} (\theta) = 
\begin{cases}
f(\theta) & \theta < \pi - 2 \alpha; \\
f(\theta) & \pi - 2\alpha < \theta < 2 \alpha; \\
0         & 2\alpha < \theta \\
\label{eq:3}
\end{cases}
\end{equation}
and
\begin{equation}
  P_{-1} (\theta) = 
\begin{cases}
0 & \theta < \pi - 2 \alpha; \\
f(\pi-\theta) & \pi - 2\alpha < \theta < 2 \alpha; \\
f(\pi-\theta) & 2\alpha < \theta, \\
\label{eq:3b}
\end{cases}
\end{equation}
where 
\begin{multline}
  f(\theta) = 1-2\cos(\alpha) \\
  - \frac{1}{\pi}\cos^{-1}\left( \frac{\cos \theta - \cos^2
      \alpha}{\sin^2\alpha}\right) \\
  + \frac{2\cos \alpha}{\pi}  \cos^{-1}\left(\frac{\cos\alpha (\cos
      \theta -1 )}{\sin \theta \sin \alpha}\right).
\label{eq:fth}
\end{multline}
The ``lost'' probability, i.e. $1-P_{+1}-P_{-1}$ corresponds to
geometries that are not detected and in general result in $N<1$.
Numerically integrating Equation (\ref{eq:2}), we can obtain a
relation between $\eta$ and the measured $c^{\rm meas}(r)$.

\subsection{Connection to gas angular momentum}

As discussed in \cite{1982Ap&SS..86..215P}, a fraction $f=0.04$ of galaxies has
gas angular momentum that is pointing in the opposite direction to the
apparent galaxy spin inferred from orientation of spiral arms. If we
momentarily distinguish between the actual gas spin correlation
function and apparent gas spin correlation function, we can write
\begin{equation}
  \eta_{\rm apparent}(r) = \left((1-f)^2+f^2\right)\eta(r) - 2 f(1-f) \eta(r),
\end{equation}
since correlation function of spins will receive a negative
contribution if exactly one spin (but not both) was randomly
reversed.  This simplifies to
\begin{equation}
\eta_{\rm apparent}(r) =   4 \left(f-\frac{1}{2} \right)^2 \eta(r).
\end{equation}
This has the expected properties. The spin correlation function will
become zero if exactly half the spin vectors are reversed, effectively
randomizing them and exactly following the primary correlation function
if all or none spins are reversed. For $f=0.96$ one gets that $\eta(r)
\sim 1.2 \eta_{\rm apparent}(r)$.

\section{Data \& Method}
\label{datamethod}

\subsection{Data}
\label{sec:data}

The basic data reduction is described in great detail in
\cite{2008arXiv0804.4483L} and \cite{2008arXiv0803.3247L}. We will
briefly summarise the data reduction in the following paragraph, but
the reader is invited to read the above papers if interested in the
details of the primary data reduction.

In the Galaxy Zoo project, a sample 893,212 galaxies were visually
classified by about 90,000 users. The sample was selected to be
sources that were targeted for SDSS spectroscopy, that is extended
sources with Petrosian magnitude $r < 17.77$. Additionally, we
included objects that were not originally targeted as such, but were
observed to be galaxies once their spectrum was taken. Where
spectroscopic redshifts are available, we find that they have the mean
redshift of $z=0.14$ and the objects with the highest redshift reach
$z\sim 0.5$.  The galaxies thus probe our local universe at
cosmological scales. Each object has been classified about $40$ times
from a simplified scheme of 6 possible classifications: an elliptical,
a clockwise spiral galaxy, an anti-clockwise spiral galaxy, an edge-on
spiral galaxy, a star / unknown object, a merger.  Various cuts
(hacking attempts, browser misconfigurations, etc.)  removed about 5\%
of our data. The data were reduced into two final catalogues based on
whether data was weighted or unweighted. In the unweighted data, each
user's classification carries an equal weight, while in the weighted
case, users weights are iteratively adjusted according to how well
each users agrees with the classifications of other users. In both
cases, the accrued classifications are further distilled into
super-clean, clean and cleanish catalogs of objects, for which we
require 95\%, 80\% and 60\% of users to agree on a given
classification.  In all cases, this is a statistically significant
detection with respect to random voting; however, the human
``systematical'' error associated with it is difficult to judge.  In
any case, we are in the limit where taking more data will not change
our sample beyond noise fluctuations as the votes are uncorrelated.
In \cite{2008arXiv0803.3247L} a bias of unknown origin toward
anti-clockwise galaxies was discovered and corrected for by adjusting
the cleanliness level for the clock-wise galaxies to a slightly lower
value. This work uses the same data and bias correction. We note,
however, that if unaccounted for, such bias would generate a constant
offset in the correlation function that cannot mimic the correlations
we are seeing in the data. After bias correction is applied, the
numbers of clockwise and anti-clockwise galaxies are the same within
Poisson noise in each sample.

\begin{figure}
  \centering
  \includegraphics[width=\linewidth]{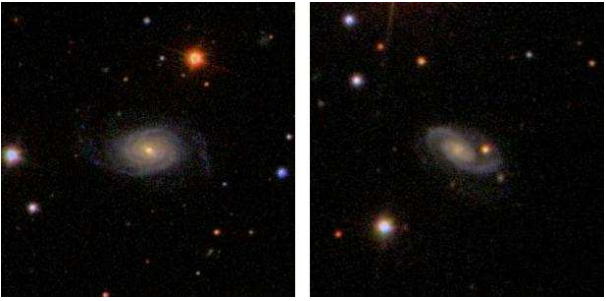}
  \caption{This figure show a pair of typical galaxies from our
    \emph{clean} catalogue. The left image is an anti-clockwise
    (S-like), while the right is a typical clockwise (Z-like) galaxy.}
  \label{fig:szfig}
\end{figure}

We decided to use the 80\% clean weighted sub-sample. We stress that
the decision to work with 80\% clean sample was made in advance and
was not chosen to maximize our signal. We show an example of a typical
clockwise and anti-clockwise spinning galaxies from a clean catalogue
in the Figure \ref{fig:szfig}.

\subsection{3D, angular and projected configuration spaces}

In the formalism of Section \ref{theory}, we have always referred to distance between
two galaxies as being $r$, the physical distance in real-space. In
practice, it can be any measure of distance between galaxies. 
In this work we use three different distance measures:
\begin{enumerate}
\item  \emph{Angular distances}. These have the advantage of producing
  the highest number of pairs. We denote the corresponding correlation
  function with $c(\theta)$.

\item \emph{Real space distance}. We use the distance in the redshift
  space for pairs of galaxies for which both spectroscopic redshifts
  are known. These are not the true 3-dimensional distances, but are
  instead distances in the red-shift space and therefore affected by
  the fingers-of-god effects \citep[see
  e.g.][]{1998ASSL..231..185H}. Since the axis of sub-halos are
  correlated with the shape of the parent halo \citep[see
  e. g.][]{2005ApJ...627..647B}, there exist correlations in the ratio
of edge-on to face-on spirals as a function of projected distances
from the centre of the halo. This considerably complicates any
correction for fingers-of-god effects and therefore we do not attempt
this correction, since effects are likely to be sudominant.  A
concordant flat cosmology with $\Omega_{\rm m}=0.25$ was assumed when
calculating distances.  We denote the corresponding correlation
function with $c(r)$.

\item \emph{Projected distances}. There distances are the transversal
  component of the distance vector connecting two galaxies with known
  redshift. If only one galaxy in the pair has a known redshift, we
  assume the other galaxy to have the same redshift.  The advantage of
  this distance measure is that it is not affected by the redshift
  space distortions and that the number of pairs is significantly
  larger than in the case of real space distances. We denote the
  corresponding correlation function with $c(p)$.
\end{enumerate}

For each of the above distance measures, we first located all galaxy
pairs in our sample, that are less than 2000 arc seconds or 3 Mpc/$h$
or 1 Mpc/$h$ projected apart. This gave us three sets which we
describe in the Table \ref{tabl:1}.

We have then removed rogue pairs. In the primary SDSS pipeline
analysis, every object is assigned an SDSS ID. Large nearby galaxies
are often associated with more than one ID, as various knots and
substructure of the galaxy are recognised as sources by the reduction
software. All such IDs are therefore classified as the same galaxy,
resulting in spurious positive correlation at the shortest
distances. Our automatic mechanism removed all pairs for which their
angular separation is less than $1.5 \max (r_p)$, where $\max(r_p)$
denotes the larger of the two Petrosian radii
\citep{1976ApJ...209L...1P}. This did remove the majority, but not all of
the rogue pairs.  Therefore, the closest pairs (at angular separations
of less than $3r_p$) in each category were examined by hand and 69
additional SDSS objects were removed.

\begin{table*}

\begin{tabular}{c|ccc}
\hline
Property & Angular & Real space & Projected \\
\hline
Number of pairs & 34031 & 8005 & 24271\\
Number of gal. & 20827   & 7979 & 25272  \\
Mean $z$  & 0.08 & 0.05 & 0.07 \\
\hline
$\Delta \chi^2$ exponential & 9.76 & 9.12 & 5.89  \\ 
Evidence ratio &    9 &    6 &    1 \\ 
$\vspace*{0.3cm} a$ & $ 0.94^{+0.39+0.54+0.56}_{-0.48-0.80-0.92} $&$ 0.35^{+0.19+0.42+0.86}_{-0.16-0.27-0.34} $&$ 0.55^{+0.60+0.89+0.95}_{-0.47-0.54-0.55} $\\
$\vspace*{0.3cm} a$ & $ 23.41^{+11.37+42.47+73.24}_{-6.35-13.04-22.41} $&$ 0.37^{+0.16+0.41+0.61}_{-0.11-0.23-0.37} $&$ 0.02^{+0.08+0.37+0.47}_{-0.01-0.02-0.02} $\\
\hline
$\Delta \chi^2$ Gaussian & 9.82 & 11.52 & 6.71  \\ 
Evidence ratio &   11 &   16 &    1 \\ 
$\vspace*{0.3cm} a$ & $ 0.60^{+0.25+0.38+0.40}_{-0.27-0.49-0.58} $&$ 0.24^{+0.10+0.22+0.41}_{-0.09-0.17-0.23} $&$ 0.44^{+0.31+0.51+0.55}_{-0.37-0.43-0.44} $\\
$\vspace*{0.3cm} b$ & $ 26.76^{+9.36+34.11+68.56}_{-6.35-13.71-25.42} $&$ 0.44^{+0.13+0.31+0.52}_{-0.10-0.20-0.42} $&$ 0.02^{+0.05+0.36+0.47}_{-0.01-0.02-0.02} $\\
\end{tabular}
\caption{This table shows the basic information about the datasets
  used in this work. When calculating the mean redshift, only subset
  of galaxies with redshift is used and we average over galaxies and
  not galaxy redshifts.  We also report values of best fit 
 $\Delta  \chi^2$, Bayesian evidences and parameters of our fits.  Note that
  evidence here is the evidence ratio and not its logarithm. }
\label{tabl:1}
\end{table*}

\subsection{Determination of $\alpha$ angle}

\begin{figure}
  \centering
  \includegraphics[width=\linewidth]{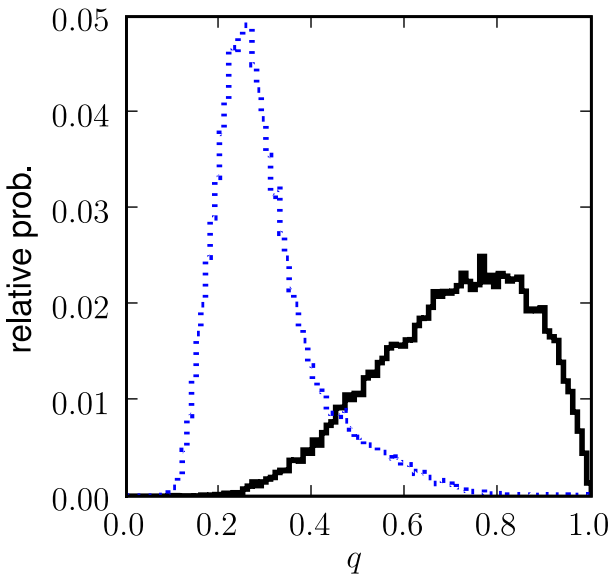}
  \caption{This figure shows the histogram of distributions of $q$
    values for galaxies classified as face-on spirals (solid black,
    classifications 2,3) and edge-on spirals (dashed blue,
    classification 4) }
  \label{fig:el}
\end{figure}

As discussed in the Section \ref{theory}, we need to estimate the
value of $\alpha$, the maximum angle of inclination at which the
spirals have a measured spin orientation rather than a being
classified as ``edge-on'' spirals. To do this, we use the adaptive
second moments \citep{2002AJ....123..583B} from the SDSS pipeline,
namely $e_\times$ and $e_+$ to calculate the axis ratio, following
\cite{Ryden:2003vd}:
\begin{equation}
  q = \left(\frac{1-e}{1+e} \right)^{1/2},
\end{equation}
where $e=\sqrt{e_\times^2+e_+^2}$.  In Figure \ref{fig:el} we show the
distribution of $q$ values for spirals galaxies classified as face-on
(of either spin orientation) and edge-on spirals. As
expected, the two populations occupy the two corners of possible
values of $q$, but there is a significant overlap. Intrinsic
ellipticities, non-zero thickness of the disk and potential
human-induced selection effects likely complicate
things. We have attempted to model intrinsic ellipticities in the
spirit of \cite{1997AJ....113...22G}, but difference was negligible. 

A plausible range of the cut-off $q$ is 0.2 - 0.5, giving the values
of $\alpha$ between $60^{\circ}$ and $80^{\circ}$. If we numerically
integrate Equation (\ref{eq:2}) as explained in Section
\ref{sec:small-scale-corr}, we get
\begin{equation}
  \eta(r) \sim m c^{\rm meas}(r),
\end{equation}
with the value of $m$ between $\sim 0.6$ and $\sim 0.9$. We will
assume a systematic bias associated with this effect to be
$m=0.75\pm0.15$.  Adding to this the effect of the random reversing of
galaxy spins and allowing a liberal 50\% enhancement of the systematic
error due to an \emph{ad-hoc} assumption in Equation (\ref{eq:bala}),
we arrive at

\begin{equation}
  \eta(r) = (0.9 \pm 0.3) c^{\rm meas}(r).
\label{eq:conv}
\end{equation}

\subsection{Correlation function measurement}

Our basic method is to measure $c(r)$ and its errors and then to infer
constraints on $\eta(r)$.

To measure $c(r)$, we note the following. For a pair of galaxies,
whose spins are $\ts_i$ and $\ts_j$, the product $\ts_i \ts_j$ can be
either +1 or -1 with probabilities $p_{\pm 1}$. Since
$p_{+1}+p_{-1}=1$ and the expectation value of $\left< \ts_i
  \ts_j\right>=p_{+1}-p_{-1}=c(r)$ it follows that

\begin{equation}
  p_{\pm 1} = \frac{1 \pm c(r)}{2}
\end{equation}

Therefore, one can write the likelihood function for $c(r)$ as

\begin{equation}
  P(c(r) | {\rm data}) \propto P({\rm data} | c(r)) = \prod_k \left(\frac{1+d_k c(r_k)}{2}\right),
\end{equation}

where index $k$ runs over all pairs of galaxies in the sample and
$d_k=\ts_i \ts_j$ is the spin product for the $k$-th pair whose
distance is $r_k$.  In practice we work with the log likelihoods
\begin{equation}
  \log P(c(r) | {\rm data}) = \sum_j \log (1+d_j c(r_j)) + {\rm const.}
\end{equation}

We use three possible forms for $c(r)$. First we assume a stepwise
shape for $c(r)$ and measure it in bins. Second, we use two
2-parameter families of curves that seem to describe our data fairly
well: an exponential
\begin{equation}
  c(r) = \min\left\{ 1,a e^{-r/b} \right\}
\end{equation}
 and a Gaussian
\begin{equation}
   c(r) = a e^{-r^2/2b^2}.
\end{equation}

This parameter space is so small that it can be efficiently explored
using grid based methods and more advanced Markov Chain methods are not
necessary.

\section{Results}
\label{sec:results}

\begin{figure}
  \centering
  \includegraphics[width=1.1\linewidth]{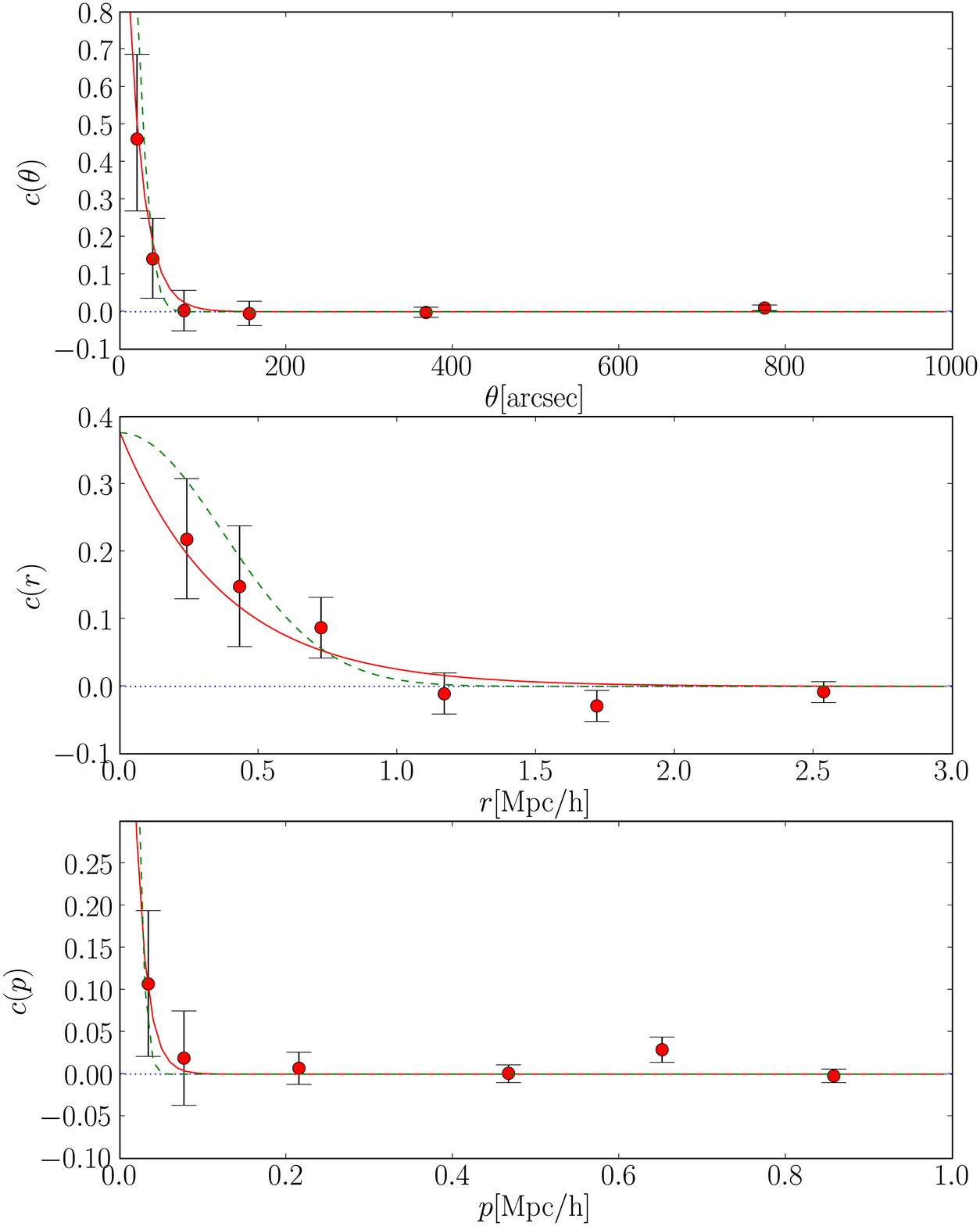}
  \caption{This figure shows the constrains on the binned correlation
    function $c$ for angular (top), redshift (middle) and projected
    (bottom) spaces. Two lines correspond to our best fit exponential
    (solid red) and Gaussian (dashed green) fits.}
  \label{fig:1}
\end{figure}

In Figure \ref{fig:1} we plot the results of our binned estimation of
$c(r)$.  From the two figures it is immediately clear that there is a
hint of an excess at low values of $r$.  The statistical significance
of this excess is marginal, at about $\Delta \chi^2$ of 7.5, 14.2 and
5.6 for angular, real and projected distances with 6 extra degrees of
freedom associated with 6 bins. This corresponds to 2-3 $\sigma$
detection in the redshift space but a non-detection in other spaces.

\begin{figure}
  \centering
  \includegraphics[width=1.1\linewidth]{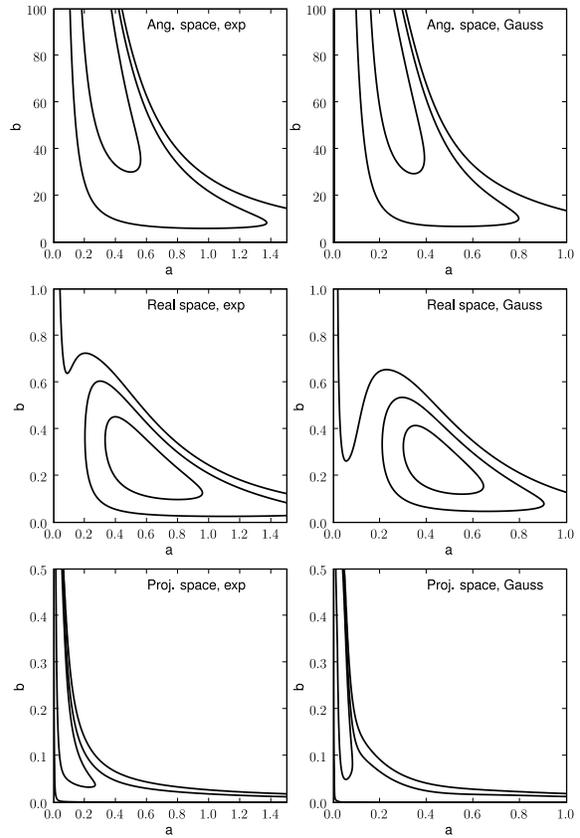}
  \caption{This figure shows the constrains on the $a$-$b$ plane for
    all datasets and models under consideration. Thick lines enclose
    68.3\%, 99.4\% and 99.7\% enclose likelihood contours for the
    weighted sample. Thin lines are the same for unweighted
    sample. The top and bottom rows show results in real and angular
    spaces respectively. The left and right columns the exponential
    and Gaussian fittings exponentially.}
  \label{fig:2}
\end{figure}

To understand this excess better, we calculate the probability
contours on the $a$-$b$ plane using exponential and Gaussian
likelihoods. These are plotted in the Figure \ref{fig:2} and the
relevant numbers are in the Table \ref{tabl:1}.  How significant are
these detections? The improvement in $\chi^2$ is between 9 and 12 with
respect to zero correlation in angular and redshift cases with two
free parameters. Within a frequentist approach this is significant at
2-3 sigma level. The excess at low redshift is not significant in the
case of projected distances, although visually the low distance points
are not incompatible with an excess.

A more appropriate statistical procedure is the Bayesian evidence
\citep{2003MNRAS.341L..29S, 2005PhRvD..71f3532B,2007MNRAS.378...72T}
which we calculate for all our 2 model parametrizations and also show
in table \ref{tabl:1}. These can be calculated exactly for a simple
problem like ours.  Evidence depends weakly on the prior size and in
this we chose the prior on $a$ between 0 and 1/1.5 for
Gaussian/exponential case and b between 0 and 1000 arc sec or 1
Mpc/$h$ or 0.5 Mpc/$h$ projected.  Regardless of the exact number
employed, the evidence ratio is between a few and a few tens units
implying a weak evidence or a hint for angular and redshift spaces,
but not for the projected space. This is consistent with results from
the frequentist approach above.

Finally, we acknowledge the fact that the exponential and Gaussian
form were chosen \emph{a-posteriori}, after seeing the data and hence
the improvements in fits contain a subjective \emph{a-posteriori}
factor.

\subsection{Systematics}

We can now briefly discuss some of the main systematic effect that
might affect our measurements.

\emph{Rogue pairs.} As discussed in Section \ref{sec:data}, we
manually looked at all pairs in the clean sample and discarded rogue
pairs. It is an important systematic check, because we have at the
same time convinced ourselves that manually classifying a small subset
(80 galaxies) of the total sample gave consistent results.

\emph{Weighting.}  Repeating our measurements with unweighted data,
changes results by less than 5\%.

\emph{Cleanliness level}. We have repeated the analysis with the
super-clean sample. There are many fewer galaxies in the super-clean
sample \citep{2008arXiv0804.4483L} and so the statistical significance decreases considerably. We
have no significant detection in any of the spaces considered. The
errorbars increase by a factor of 2 to 2.5, but the central values in 
individual bins remain consistent. While the statistical power is
decreased, the final signal is consistent with the results presented
above.

\begin{figure}
  \centering
  \includegraphics[width=1.1\linewidth]{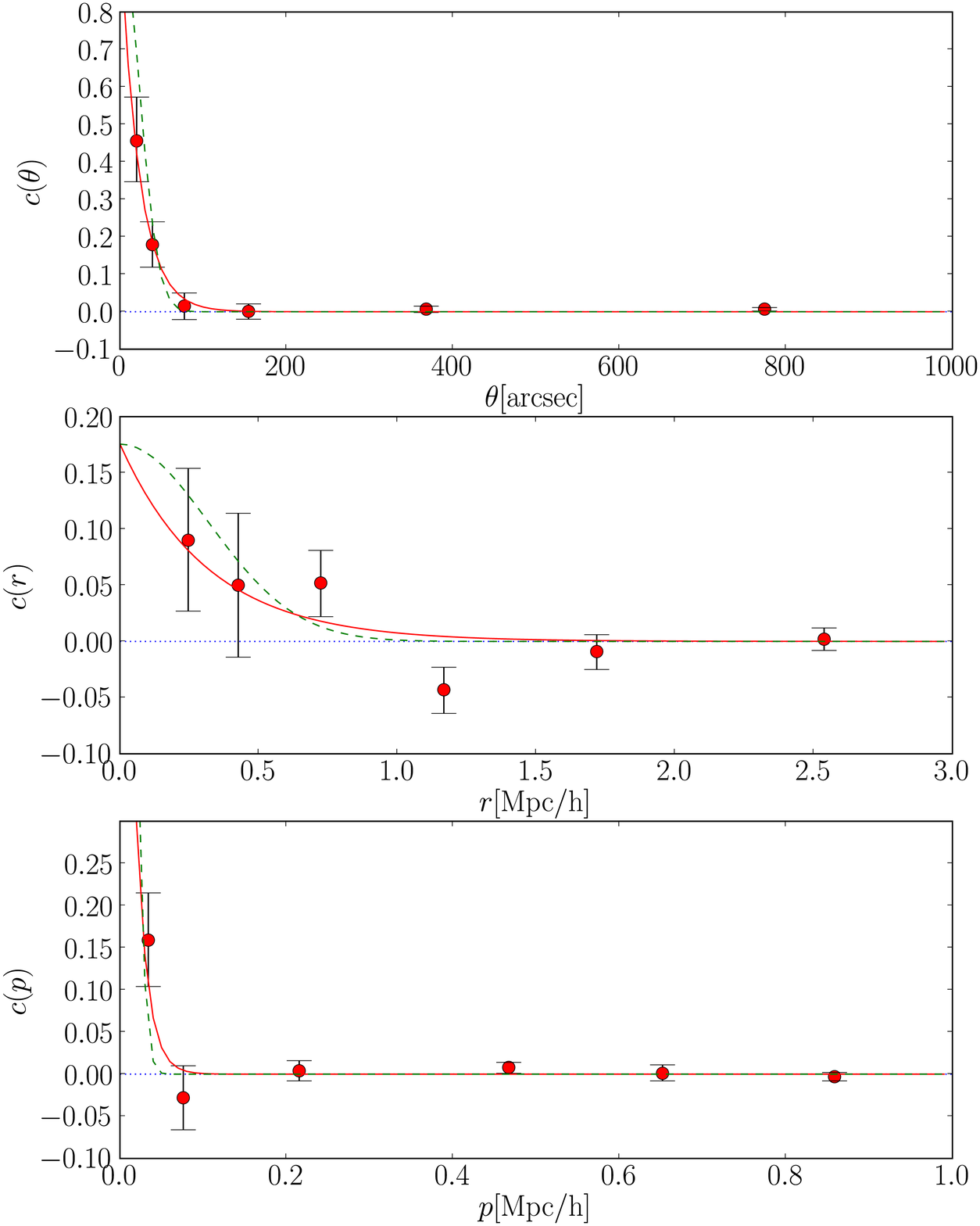}
  \caption{Same as Figure \ref{fig:1}, but for \emph{cleanish}
    sample. These results are likely to be affected by the rogue pairs
    - see text for discussion.}
  \label{fig:60}
\end{figure}

We have also repeated our measurements with the the \emph{cleanish}
sample that requires 60\% of votes to agree. The results imply strong
detections in both angular and projected samples, but with a lower
significance in redshift space. We show their results in the Figure
\ref{fig:60}. The high-confidence with which the results are detected
in angular and projected spaces is likely to be deceiving, as the
rogue pairs have not been manually cleaned for these samples. The
decrease of signal in the redshift space indicates that the signal is
indeed getting lower due to noise introduced by low significance.

\emph{Selection effects.} Another important question is whether there
is any physical difference between our redshift sample and angular
sample and how do these samples compare to the general SDSS sample.
To do this we have divided the galaxies that formed our pairs at
closest distances into those for which we have redshift information
and those for which redshift information is not available.  When
comparing colors and magnitudes we find that there is no evidence that
objects with and without redshifts are drawn from different magnitude,
$u-r$ colour or petro radius size distributions. The reason for some
objects lacking redshift is therefore probably incompleteness due to
fibre collisions.  We therefore find no evidence that the correlation
in the angular sample is of a different physical origin than the
correlation in the real-space sample.

\section{Discussion and Conclusions}
\label{sec:end}

We are now in position to make a synthesis of our results.  The
redshift-space results show that there is a significant correlation of
$c(r)\sim 0.15$ in the projected galaxy spins up to the $\sim$ 0.5
$h$/Mpc. The angular correlation, shows larger correlations of
$c(\theta) \sim 0.4$ that are significant correlations up to 30
arc-seconds, which roughly corresponds to projected distances of about
0.03 Mpc/$h$, since mean redshift of $0.08$ correspond to distance of
$\sim 230$ Mpc/$h$. This is consistent with the redshift-sample
results - both exponential and Gaussian fits do predict $c(r)$ to
raise to $\sim$ 0.3--0.4 as $r$ goes to zero. A consistent picture is
therefore the following: The angular sample detect correlations at the
shortest distances, where majority of pairs are physical associations,
but these get diluted at larger distances due to interlopers. The
redshift-space correlations track these correlations to larger
physical distances. Projected space pairs do not have enough
signal-to-noise to detect these correlations at high significance.

How do these results compare to theoretical predictions? Simple models
as those suggested in \cite{2000ApJ...543L.107P} (equations
(\ref{eq:pry}) and (\ref{eq:prx})) predict a vanishing $\eta$ and
hence we have directly detected a non-random distribution of inertia
tensors. Within the standard model, the reason for correlations of
moments of intertia are the correlations of these with the (slowly
varying) tidal field. On the other hand, if moments of intertia are
\emph{perfectly} aligned with the tidal field, the tidal torque cannot
produce any angular momentum and therefore the resulting angular
momentum is due to the residual 10\% of misaligment
\citep{Porciani:2001er}. The stunning outcome of our result, if
confirmed, is that even these 10\% misaligments are correlated from
(sub-)halo to (sub-)halo.

What is also interesting is, however, that in \cite{Porciani:2001db}
$\eta$ correlations have not been detected in simulations at $z=0$ at
all separations. In particular, $\eta<0.02$ at $r=0.5$ Mpc/$h$. A
virtually identical result has been found by
\cite{2005ApJ...627..647B}, who also find $\eta<0.02$ at
$r=0.5$Mpc/$h$ (our $\eta$ is their $\xi_{LL}$). This is in tension
with our results even after conversion factors in Equation
(\ref{eq:conv}) are taken into account. There are many reasons that
explain why our results are not directly comparable to the above
work. First, they are comparing individual dark matter halos. In our
case, we see the signal at pair separations of less than 1 Mpc. At
such distances, one-halo pairs (pairs of galaxies that reside in the
same dark matter halo), dominate over two halo pairs (pairs in which
two galaxies occupy two different halos).  By selecting spiral
galaxies, we are essentially selecting pairs that are composed of
satellites residing in the same halo, rather than pairs compromised of
central halo galaxies. The latter are bright ellipticals and hence
inaccessible using our method. Unfortunately, not very much
theoretical work has been done for spin correlations of
substructure. The most relevant paper in the literature is
\cite{2005ApJ...629L...5L}, which, however, still uses the chirality
agnostic model of \cite{2000ApJ...543L.107P} and does not calculate
the chiral correlation function. More work on the theoretical side and
$N$-body side is required to understand the implications of our
results. Hopefully, the results could be turned around and help us
understand what kind of substructure spiral galaxies occupy in a
typical dark matter halo.

It is therefore imperative that our observational protocol is
simulated on a large enough $N$-body simulation, for example the
\emph{Millennium Simulation} \citep{Springel:2005nw} or
\emph{MareNostrum Universe} \citep{2007ApJ...664..117G}
simulations. There, halos and sub-halos hosting spiral galaxies can be
identified and those, whose inclination with respect to a given
observer is small enough to be considered face-on, should be
correlated. This would  result in a quantity $c(r)$ that is directly
comparable to the observables that we constrain with the Galaxy Zoo data.

Another interesting aspect of our results is that, for spiral
galaxies, we essentially exclude large and random misaligments between
gas and dark matter angular momenta. Since the dark matter is
dynamically dominant, gas angular momenta can only be correlated if
they are so due to correlations between dark matter.

Finally, it is tempting to combine our measurements with the
ellipticity measurements to improve signal-to-noise and remove some
systematic. Note, however, that our 1-bit signal divides a four-fold
degeneracy into a two-fold one and thus this is a non-trivial task
which will be left for the future.

To conclude: We have tentatively detected a chiral correlation
function in the spins of spiral galaxies. This correlation function
vanishes in the simplest models based on tidal torque theory. Our
results indicate, that moments of intertia of protohalos that end up
hosting spiral galaxies are correlated at distances less than $\sim$
0.5 Mpc/$h$.These short distances imply that these protohalos are
often likely to be substructures of massive halos.  More work is
required to understand these results at a quantitative level.


\section*{Acknowledgements}
 We would be nowhere without the amazing contributions from all 
the GZ members, forum contributors, and other essential volunteers!

AS thanks Cristiano Porciani for bringing up the important distinction
between one and two-halo pairs and Uro\v{s} Seljak for useful
literature tips. 

AS is supported by the inaugural BCCP Fellowship. CJL
acknowledges support from an STFC Science in Society Fellowship.


\appendix

\section{Calculations of ``seen pair'' probabilities. }
\label{apx}
The question that we want answer is: For a given pair of galaxies,
whose spin vectors are at angle $\cos \theta=\mu$, what is the
probability of an observer seeing the pair with the same sense of
galaxy rotation, what opposite senses of galaxy rotation and not
seeing them at all due to selection effects?

If $\alpha=\pi/2$, the result can be obtained by considering each spin
in turn. The first spin divides the unit sphere of possible observer
directions into two half-spheres, depending on the sign of its
projected spin. When two spins are considered, the intersection of the
two half-spheres are two lunes. The thickness of the lune of opposite
spins is $\theta/\pi$, leading to the result in Equations
(\ref{eq:plmin2}) and (\ref{eq:plmin}).

When $\alpha<\pi/2$ the dividing line between the two half-sphere
becomes a band of angular thickness $2(\pi/2-\alpha)$ and the two
half-spheres shrink to two spherical caps of radius $\alpha$. The
overlapping area of the spherical caps separated by $\theta$ is $4\pi
f(\theta)$, where $f(\theta)$ is given by Equation (\ref{eq:fth})
\citep{1230111}.

If $\alpha>\pi/4$, the intersection of the 4 cups give four ``trimmed''
lunes. There are three possibilities.
\begin{itemize}
\item $\theta<2(\pi/2-\alpha)$. Both spins are in roughly same
  direction and the opposite spin ``trimmed'' lunes are squeezed to
  zero area. Hence $P_{+1} = f(\theta)$ and  $P_{-1}= 0$

\item $2(\pi/2-\alpha)<\theta<2\alpha$. General situation in which all
  four ``trimmed'' lunes have finite area. We have $P_{+1} =
  f(\theta)$ and $P_{-1}= f(\pi-\theta)$

\item $\theta>2\alpha$. Both spins are in roughly opposite
  directions and opposite spin ``trimmed'' lunes are squeezed to zero
  area. In this case$P_{+1} = 0 $ and  $P_{-1}= f(\pi-\theta)$ 
\end{itemize}

These results imply Equations (\ref{eq:3}) and (\ref{eq:3b}). We have
tested these analytical predictions using a Monte Carlo code. 

\bibliographystyle{mn2e}
\bibliography{cosmo,cosmo_preprints}
\bsp

\label{lastpage}

\end{document}